\documentstyle[11pt,epsfig,a4,rotating]{article}
\begin{document}

\def\Journal#1#2#3#4{{#1} {\bf #2}, #3 (#4)}

\def\etal{{\it et\ al.}}
\def\NCA{\em Nuovo Cim.}
\def\NIM{\em Nucl. Instrum. Methods}
\def\NIMA{{\em Nucl. Instrum. Methods} A}
\def\NPB{{\em Nucl. Phys.} B}
\def\PLB{{\em Phys. Lett.}  B}
\def\PRL{\em Phys. Rev. Lett.}
\def\PRC{{\em Phys. Rev.} C}
\def\PRD{{\em Phys. Rev.} D}
\def\ZPC{{\em Z. Phys.} C}
\def\ASP{{\em Astrop. Phys.}}
\def\JETP{{\em JETP Lett.\ }}

\def\numunue{\nu_\mu\rightarrow\nu_e}
\def\numunutau{\nu_\mu\rightarrow\nu_\tau}
\def\nuebar{\bar\nu_e}
\def\numubar{\bar\nu_\mu}
\def\numubarnuebar{\bar\nu_\mu\rightarrow\bar\nu_e}
\def\nuebarnumubar{\bar\nu_e\rightarrow\bar\nu_\mu}

\title{A medium baseline search for $\numunue$ oscillations \\
at a $\nu$ beam from muon decays}
\author{A. Bueno, M. Campanelli, A. Rubbia\footnote{On leave from CERN, Geneva, Switzerland.}\\
{\it Institut f\"{u}r Teilchenphysik} \\ 
{\it ETHZ, CH-8093 Z\"{u}rich, Switzerland}}
\maketitle
\abstract{The accurate knowledge of the $\bar\nu_e$ ($\nu_\mu$)
beam produced in $\mu^-$
decays and the absence of $\nu_e$ ($\bar\nu_\mu$) contamination, 
make a future 
muon storage ring the ideal place to look for $\numunue$ ($\numubarnuebar$)
oscillations. Using a detector 
capable of electron and muon identification with charge discrimination 
(e.g., the presently running NOMAD experiment),
good sensitivities to $\numunue$ ($\numubarnuebar$)
oscillations could be achieved.
With the CERN-PS as a proton driver for a muon storage ring
of the kind envisaged for a $\mu$-collider, the LSND claim
would be confirmed or disproved in a few
years of running.}

\section{Introduction}
In a recent study\cite{longbase}, we have investigated the possibility of
performing a long baseline experiment using a neutrino
beam from muon decays. In this paper, we concentrate
on the possibility of a medium baseline experiment, motivated
by the outstanding problem of the verification of the Los Alamos 
LSND experiment, which claims evidence for neutrino oscillations in the channels 
$\numubarnuebar$\cite{DAR-LSND} and $\numunue$\cite{DIF-LSND}.
\par
The latest KARMEN results\cite{karmen} have come very close to contradicting the
LSND claim, however the experimental sensitivity is marginal to 
conclusively exclude or confirm completely the LSND allowed solution.
\par
The probability for neutrino oscillations between two families is given by 
\begin{eqnarray}
P(E_\nu)=\sin^2 2\theta \sin^2(1.27 \Delta m^2 \frac{L}{E_\nu})
\label{eq:pro1}
\end{eqnarray}
where $\theta$ is the mixing angle between the two neutrino flavours, and
$\Delta m^2$ (in $eV^2$) is the difference of the squares of the neutrino masses.
\par
The part of the LSND experiment solution with $\Delta m^2>\approx 10\ eV^2$ has recently 
been shown to be inconsistent with the data taken with the NOMAD detector\cite{nomad}
at the present CERN-WANF neutrino beam.
In the NOMAD detector where electron identification capability is 
rather good and well understood, the 
current search for $\numunue$ oscillations\cite{slava} is limited in sensitivity by the
systematic uncertainty of the prediction of the intrinsic $\nu_e$
component of the beam originating from primary $K^+$, $K^0_L$ and $\mu^+$ decays.
\par
New experiments have been proposed to cover the $\Delta m^2-\sin^22\theta$ 
regions indicated by 
the LSND claim: 
\begin{itemize}
\item ICARUS at Jura\cite{jura} would use
the currently existing CERN WANF neutrino beam, extending
the baseline to 17 km (allowing also tau appearance). 
\item At Fermilab, the MiniBOONE\cite{boone} experiment
would address the low $\Delta m^2$ region by 
using a neutrino beam from the 
FNAL booster. The average energy is 1.5 GeV and the distance from the source to the 
detector is 1 km.
\item The CERN/I216 letter of intent\cite{cernps} suggested to reuse the old CERN-PS 
neutrino beam in order to obtain a low energy beam. 
\end{itemize}
\par
So far, MiniBOONE is the only approved experiment.  
To improve the oscillation result at low $\Delta m^2$, the proponents of MiniBOONE
have chosen an optimized $L/E_\nu$ in order 
to maximize the oscillation signal compared to backgrounds. 
This experiment should provide a strong signal in case the LSND effect is 
really 
due to neutrino oscillations. It suffers however from irreducible backgrounds 
associated
to the intrinsic $\nu_e$ component of the beam and from misidentified
electrons\cite{boone}, that could make more problematic an
interpretation of the observed effects.
\par
We suggest that a complementary approach is to try to 
reduce background sources as much as possible, in order 
to obtain good statistical sensitivities, even in the
case of small oscillation probabilities.
We take advantage from the fact that when a neutrino beam is 
produced from muons of a definite sign, 
for example from the decay of negative muons\footnote{In this paper we always refer to the 
decay of negative muons. The same considerations are obviously valid in the
case of positive muons.}, only one kind
of flavor-antiflavor is produced, i.e.
\begin{eqnarray}
\mu^- \rightarrow e^-\bar\nu_e\nu_\mu.
\end{eqnarray}
\begin{itemize}
\item In case of $\numunue$ oscillations, there will be appearance of $\nu_e$ neutrinos while 
the unoscillated beam contained only $\bar\nu_e$'s. Charge discrimination in the detector will 
trivially separate the two types of neutrino components:
\begin{eqnarray}
\nuebar + N & \rightarrow & e^++X\ \ \ \ \ \rm   intrinsic\ beam\ component, \\
\nu_e +N & \rightarrow & e^- + X\ \ \ \ \ \rm  oscillated
\end{eqnarray}
\item In case of $\bar\nu_e\rightarrow\bar\nu_\mu$ oscillations, 
there will be appearance of $\bar\nu_\mu$ neutrinos while the
unoscillated beam contained only $\nu_\mu$'s. 
\begin{eqnarray}
\nu_\mu +N & \rightarrow & \mu^- + X\ \ \ \ \ \rm   intrinsic\ beam\ component,\\
\numubar + N & \rightarrow & \mu^++X\ \ \ \ \ \rm   oscillated
\end{eqnarray}
\end{itemize}
We stress that this experiment would test
the transition $\numunue$ and its CP-conjugate $\bar\nu_\mu\rightarrow\bar\nu_e$
at the same time.
\par

\section{Event rates}
We have already discussed our assumptions concerning the muon source in
our previous paper\cite{longbase}
and recall here only the important points.
Based on muon collider studies\cite{muonco}, we assume that the neutrino beam 
will be produced by the decay of a large quantity of stored muons\cite{geer}. 
The muons are produced in the 
decay chain of pions behind an appropriate target and are subsequently captured, 
cooled, accelerated and stored into a ring where they are let to decay. 
We compute the total number
of muons accelerated in the following way:
\[N_{\mu}=f_{bunch}\times N_{p/bunch}\times Y_{\pi/p}\times Y_{\mu/\pi}\times
t\]
where $f_{bunch}$ is the bunch repetition rate, $N_{p/bunch}$ is the number
of protons per bunch, $Y_{\pi/p}$ is the yield of pions per proton,
and $Y_{\mu/\pi}$ is the yield of muons per pion.
\par
We need a proton driver in order to produce high energy protons
for the muon source.
The currently existing CERN-PS machine is able to accelerate\footnote{See
the discussion in Ref. \cite{cernps}
and we also stress that this number depends on a shared/dedicated mode of 
operation. We assume that neither LEP, nor LHC will be running during
this time.}
$f_{bunch}\times N_{p/bunch}\times t \approx 2.5\times 10^{20}$ protons in
a year at an energy of 19.2 GeV.
\par
We assume that $Y_{\pi/p}\times Y_{\mu/\pi} = 0.08$ muon per proton 
will be trapped and accelerated into
the storage ring\cite{muonco}.
With this assumption,
the total number of muons accelerated is $2\times 10^{19} \mu/year$.
We consider that after 2 years of running, a total of $4\times 10^{19}
\mu$ will have decayed inside the storage ring.
With an optimized geometrical configuration of the storage ring
like a ''cigare'', in which two long straight sections are closed
by two small arcs where the muons are strongly bent, 
half the muons will produce 
neutrinos in the right direction. 
\par
\begin{figure}
\begin{center}\mbox{
\epsfig{file=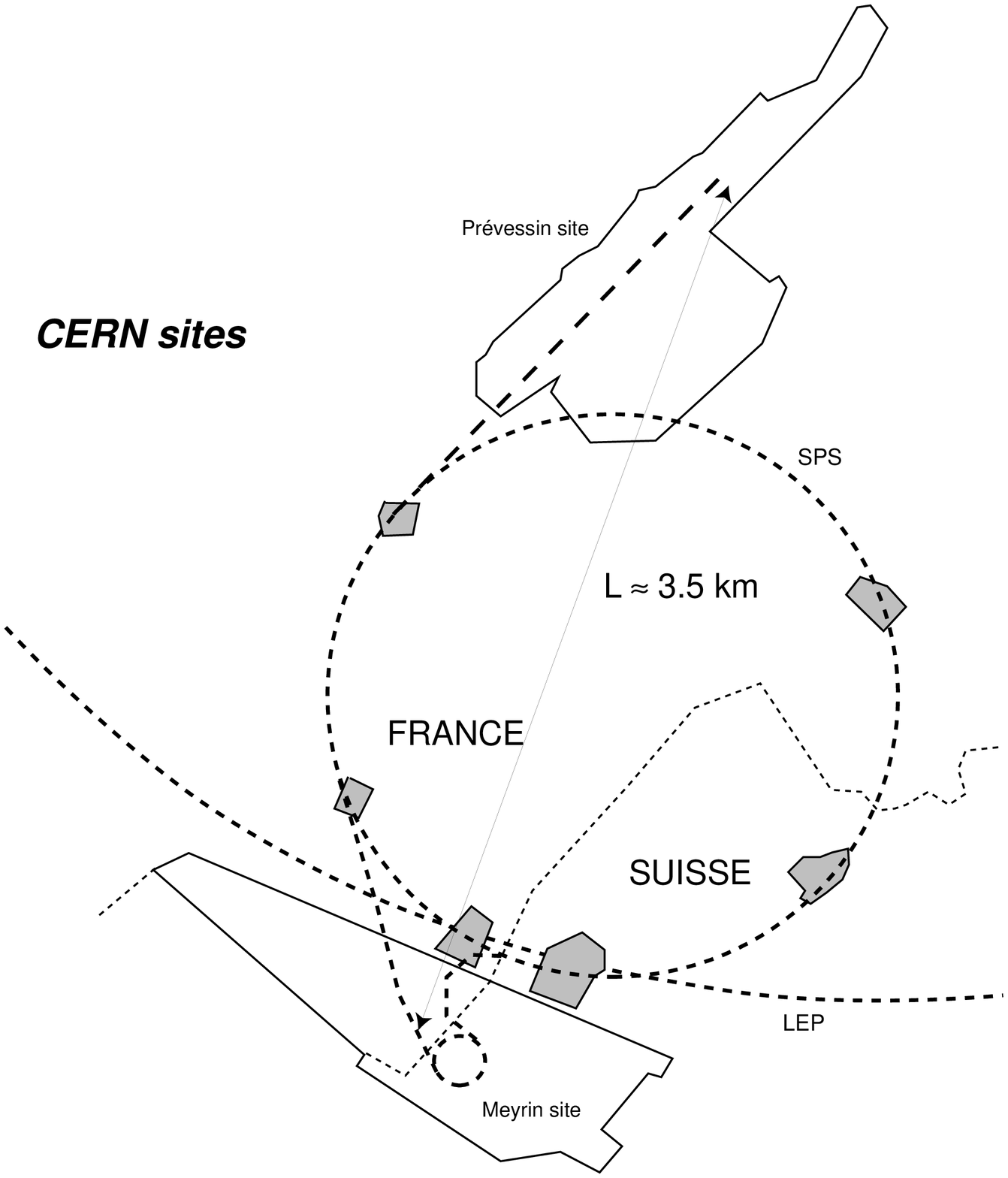,width=.9\linewidth}
}
\end{center}
\caption{
Schematic overview of the CERN sites.}
\normalsize
\label{fig:cernsite}
\end{figure}

We chose a baseline from the CERN-PS to the CERN-Prevessin 
laboratory, close to the North Area. The distance from the PS is
about 3.5 km (see Figure \ref{fig:cernsite}).
We tentatively use the existing NOMAD detector as a target (total mass of 2.6 tons). 
It is presently located in the West Area and we assume that 
it can be moved to the North Area. 
\par
\begin{figure}
\begin{center}\mbox{
\epsfig{file=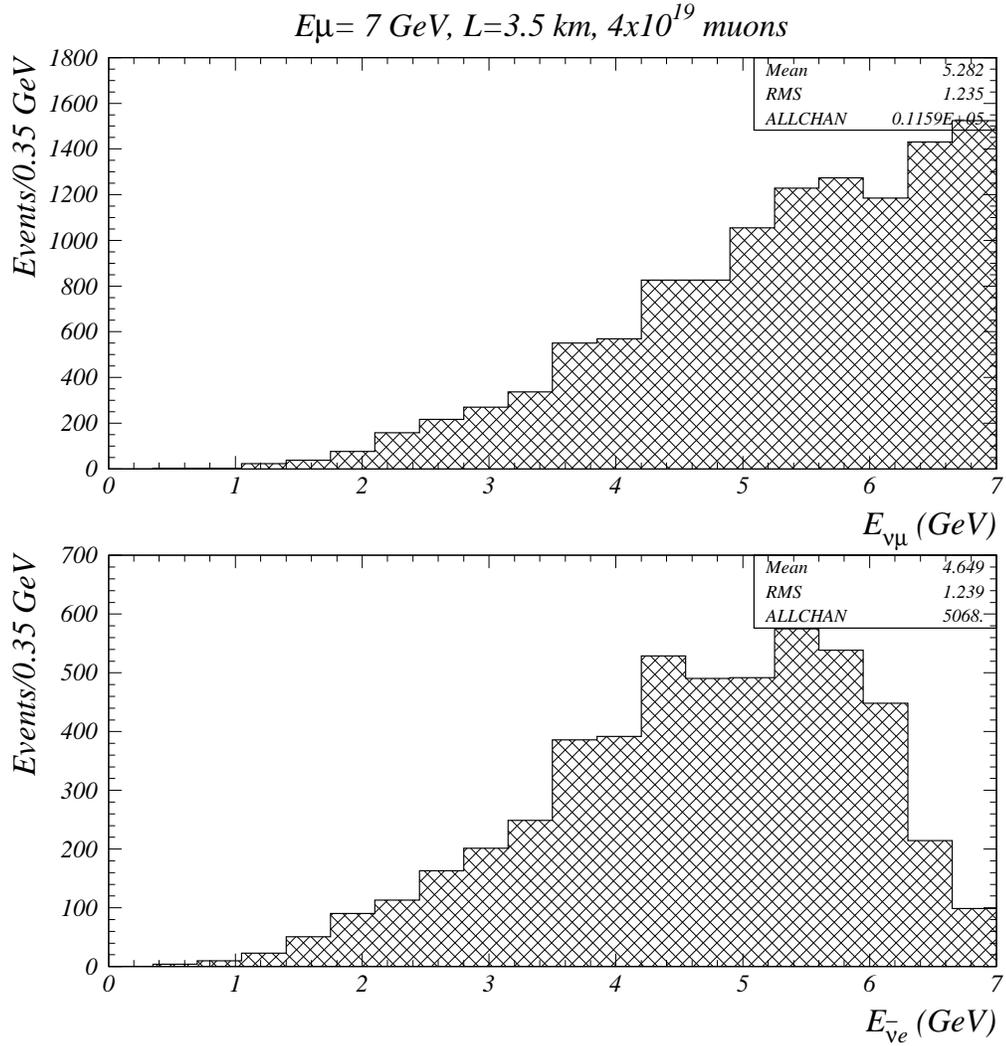,width=.9\linewidth}
}
\end{center}
\vspace*{-0.3cm}
\caption{
Predicted muon neutrino and electron anti-neutrino charged current event energy
distribution for an integrated statistics of $4\times 10^{19}$ muons.
}
\normalsize
\label{fig:oscmu}
\end{figure}

The muon beam energy is
set to 7 GeV in order to obtain a relatively low energy neutrino beam.
We compute the event rates assuming an unpolarized muon beam. 
The event rates for an integrated 
intensity of $4\times 10^{19}\mu$'s are shown in Table 
\ref{tab:rates1}.
The energy spectra of $\nu_\mu$ and $\bar\nu_e$
charged current events are shown in Figure \ref{fig:oscmu}.
For comparison, we give the expected number of oscillated
events for full mixing and various relevant $\Delta m^2$
parameters.

\begin{table}[ht]
\begin{center}
\begin{tabular}{lc}
\hline
{\it No oscillations} \\
\ \ \ \ $\nu_\mu$ CC events & 11600 $\mu^-$\\
\ \ \ \ $\bar\nu_e$ CC events & 5070 $e^+$ \\
{\it With oscillations $\Delta m^2 = 1 eV^2$, $\sin^22\theta=1$} \\
\ \ \ \ $\nu_e$ CC events & 6600 $e^-$ \\
\ \ \ \ $\bar\nu_\mu$ CC events & 3300 $\mu^+$ \\

{\it With oscillations $\Delta m^2 = 0.4 eV^2$, $\sin^22\theta=1$} \\
\ \ \ \ $\nu_e$ CC events & 1580 $e^-$ \\
\ \ \ \ $\bar\nu_\mu$ CC events & 910 $\mu^+$ \\

{\it With oscillations $\Delta m^2 = 0.2 eV^2$, $\sin^22\theta=1$} \\
\ \ \ \ $\nu_e$ CC events & 430 $e^-$ \\
\ \ \ \ $\bar\nu_\mu$ CC events & 260 $\mu^+$ \\
\hline
\end{tabular}
\caption{Neutrino event rates assuming the NOMAD detector
(2.6 tons) as a target. The muon beam has an energy
$E_\mu=7{\rm\ GeV}$ and the integrated intensity is
$4\times 10^{19}\mu$'s. The baseline is $L=3.5{\rm\ km}$.
}
\label{tab:rates1}
\end{center}
\end{table}

\section{Results with the NOMAD detector}

We have studied the possibility to perform $\nuebarnumubar$ and $\nu_\mu\rightarrow\nu_e$
oscillation appearance searches. The event selection is based on the identification
of leading final state electrons or muons, and looks for opposite charges than
the one expected in the unoscillated beam components.
\par
We carried out the study of signal efficiency and background
estimations using a full simulation of the events and the reconstruction packages
currently used in NOMAD. All selection cuts are applied on the
reconstructed quantities.
\par
For the electron final state, the following backgrounds are the dominant ones:
\begin{itemize}
\item Charge confusion: $\nu_e + N \rightarrow e^+ +X\rightarrow$ ``$e^-$''$+X$ 
\item Asymmetrically reconstructed conversions or Dalitz decays:
$\nu_\ell + N \rightarrow \nu_\ell +X$ and $X\rightarrow e^-(e^+) +X'$
\item Misidentified hadrons: $\nu_\ell + N \rightarrow \nu_\ell +X$ and $h^-\rightarrow$ ``$e^-$''.
\end{itemize}
\par
The NOMAD detector is well suited to search for electron
charged current events. The electron identification is good and well understood. 
Charge separation of leading 
electron is also very well known. Charge confusion arises primarily from early 
bremsstrahlung with leading photon converting asymmetrically into a e$^+$e$^-$ pair where the 
e$^-$ is lost. This process is rare and the simulation gives in the low 
energy region a contamination at the level of one in ten thousand.
\par
An important source of background for the electron search 
comes from Dalitz decays and $\pi^0$ 
conversions (where the $\gamma$ converts close to the primary interaction 
vertex). Those non-prompt electrons, embedded in the recoiling hadronic 
system, are in general less isolated than the prompt ones. Hence, an 
efficient rejection of this background is obtained by demanding that the 
electron candidate has the highest transverse momentum in the event. 
Additional discrimination is achieved by means of a lower cut on the 
measured electron momentum. Remnant backgrounds, due to hadron 
misidentification, can be reduced to a negligible level by comparing the 
reconstructed electron momentum (tracking) to the associated measured energy 
(calorimetry). 

\begin{table}[tb]
\begin{center} 
\begin{tabular}{|c|c|c|c|c|c|}\hline
\multicolumn{6}{|c|}{$\nu_\mu\rightarrow\nu_e$}\\
\hline 
 & $\nu_e$ CC (\%) & $\nu_\mu$ CC & $\nu_\mu$ NC & $\bar{\nu}_e$ CC & $\bar{\nu}_e$ NC \\ \hline
Initial & 100 & 11600 & 3867 & 5070 & 2028 \\
Candidate highest $P_T$ & 59 & 196 & 99 & 1922 & 35.2 \\
$P_{ele} > 2.5 GeV$ & 41 & 4 & 2.1 & 1503 & 0.4 \\
Calorimetry & 35 & 0 & 0.7 & 1360 & 0.4 \\
Charge & 35 & 0 & $<0.2$ & $<0.2$ & $<0.1$ \\ \hline
\end{tabular} 
\end{center}
\caption{Summary of detection efficiencies and expected backgrounds for 
a $\numunue$ oscillation search.}
\label{tab:nue}
\end{table}  

\begin{table}[tb]
\begin{center} 
\begin{tabular}{|c|c|c|c|c|c|}\hline
\multicolumn{6}{|c|}{$\bar\nu_e\rightarrow\bar\nu_\mu$}\\
\hline
 & $\bar{\nu}_\mu$ CC (\%) & $\nu_\mu$ CC & $\nu_\mu$ NC & $\bar{\nu}_e$ CC & 
$\bar{\nu}_e$ NC \\ \hline
Initial & 100 & 11600 & 3867 & 5070 & 2028 \\
Candidate highest $P_T$ & 65 & 3874 & 2.5 & 0.5 & 0.4 \\
Charge & 65 & 2 & 1.1 & 0 & 0.18 \\
Isolation & 50 & $<0.1$ & $<0.2$ & 0 & $<0.1$ \\ \hline
\end{tabular} 
\end{center}  
\caption{Summary of detection efficiencies and expected backgrounds for 
a $\nuebarnumubar$ oscillation search.}
\label{tab:numubar}
\end{table} 

\par 
Table \ref{tab:nue} summarizes the results obtained for electrons, when
normalized to $4\times 10^{19}$ muons.
For an overall 35\% electron detection efficiency ($\epsilon_{e^-}$), the expected 
background is well below one event. 
\par
For the muon search, we use the standard muon identification algorithm
used in the NOMAD experiment. It has a momentum threshold of about 3 GeV
imposed by the requirement of reachability of the muon chambers.
After the cuts listed in Table \ref{tab:numubar}, 
the detection efficiency for $\mu^+$ ($\epsilon_{\mu^+}$) is 50 \%.
The background contamination for the $\bar{\nu}_e\rightarrow\bar\nu_\mu$
oscillation search is negligible. The small punch-through background
is completely suppressed by a kinematical isolation requirement, since
a misidentified muon will come from the jet.
\par
\begin{figure}
\begin{center}
\mbox{
\hspace*{-8cm}
\epsfig{file=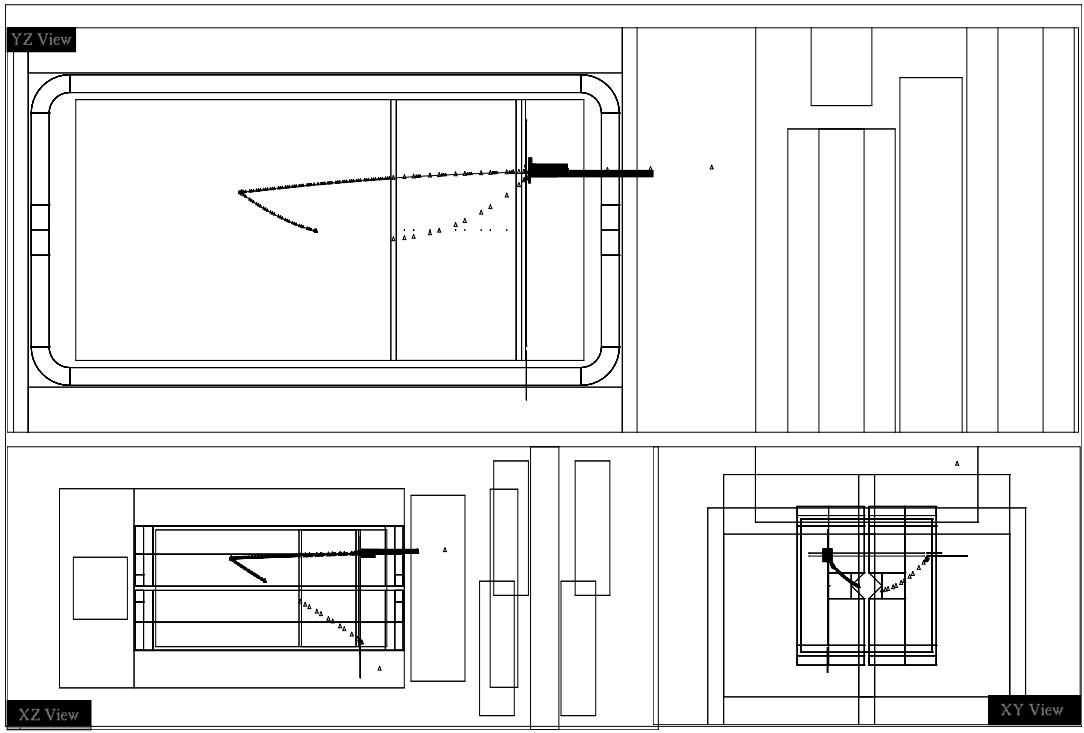,height=15.cm,angle=-90}
}
\end{center}
\caption{A $\nu_e$ CC oscillated signal event in NOMAD. The leading particle 
is 5 GeV electron stopping in the Electromagnetic Calorimeter}
\label{fig:signal1}
\end{figure}

In case of positive signal, the simultaneous use of
both channels provides a test of CP conservation in the leptonic system. 
Figure \ref{fig:signal1} shows a candidate event
for the $\numunue$ oscillation. The signal events appear as 
remarkably clean and simple to reconstruct.
\par
\begin{figure}
\begin{center}\mbox{
\epsfig{file=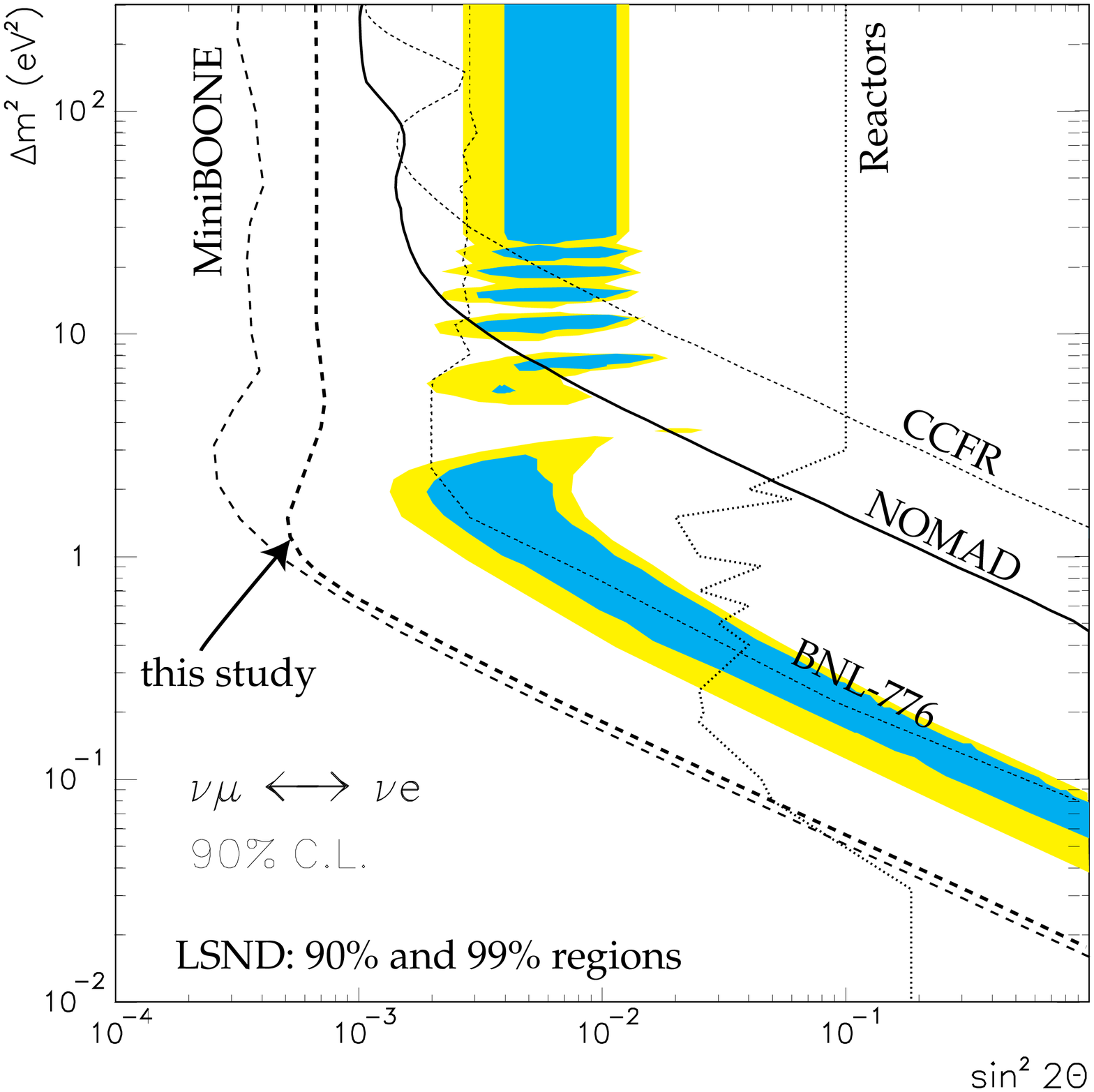,width=.9\linewidth}
}
\end{center}
\vspace*{-0.3cm}
\caption{Predicted 90\%C.L. limit for this experiment compared to expected
MiniBOONE limit for neutrino
oscillation appearance. Negative results from previous NOMAD, CCFR
and reactor experiments are also shown.}
\normalsize
\label{fig:excl}
\end{figure}

In case of negative result, combining both channels assuming CP conservation,
the 90\%C.L. limit on $\numunue$ oscillations,
for a total flux of $4 \times 10^{19}$ muons, can be obtained from the following:
\begin{equation}
P(\nu_\mu\rightarrow\nu_e) < \frac{2.3}{11600\times \epsilon_{e^-} + 
5070 \times \epsilon_{\mu^+}} = 3.5\times 10^{-4}
\end{equation}
This corresponds to a limit on the mixing angle for large $\Delta m^2$ of
\begin{equation}
\sin^22\theta < 7\times 10^{-4}\ \ \ \ 90\% C.L.
\end{equation}
and $\Delta m^2 > 2 \times 10^{-2}\ eV^2$ for maximal mixing. 
The 90\% C.L. exclusion contour is shown in Figure \ref{fig:excl}
compared to the LSND positive solution, the current limits
from NOMAD and CCFR, and the expected limit from MiniBOONE.

\section{Conclusions}

Testing the muon yield per proton
sent on a target, is the first step in the R\&D program
towards the construction of a future muon collider.
At CERN, the CERN-PS would be the natural proton source for this device. We have shown
that with this setup, and with the possibility of recycling parts of an
already existing detector (NOMAD), it would be possible to perform
an experiment to increase the
present limits on $\nu_\mu-\nu_e$ oscillations.
The area in the parameter space indicated
by the LSND result can be fully covered, with a sensitivity compared to that
of the MiniBOONE experiment, but with a completely different approach.

\newpage

\newpage
\begin{thebibliography}{000}

\bibitem{longbase} A. Bueno, M. Campanelli and A. Rubbia, hep-ph/9808485, August 1998.

\bibitem{DAR-LSND} 
C.\ Athanassopoulos \etal, (LSND Collaboration), \Journal{\PRC}{54}{2685}{1996}; 
C.\ Athanassopoulos \etal, (LSND Collaboration), \Journal{\PRL}{77}{3082}{1996};
      \Journal{\PRL}{75}{2650}{1995}. Updated results including 1997
data can be found at 
{\it http://www.neutrino.lanl.gov/LSND/papers.html}.

\bibitem{DIF-LSND} C.\ Athanassopoulos \etal, (LSND Collaboration), 
LA-UR-97-1998/UCRHEP-E191, submitted to Phys.\ Rev.\ C.;
C. Athanassopoulos \etal, \Journal{\PRL}{81}{1774}{1998}.

\bibitem{karmen} B. Zeitnitz, KARMEN Collab., proceedings of the
Neutrino98 Conference, Takayama, Japan, June 1998.

\bibitem{nomad} NOMAD Collab, J. Altegoer {\it et al.}, {\it Nucl. Instr. 
and Meth. A 404 (1998) 96}.\\
A. Rubbia, Nucl. Phys. B(Proc. Suppl.) 40 (1995) 93.

\bibitem{slava} V. Valuev, NOMAD Collaboration, proceedings of the 
International Conference on High Energy Physics, Jerusalem, Israel, 19-26 
August 1997, to be published.

V. Valuev, 
{\it ``Recherches des oscillations de neutrinos 
$\nu_\mu \rightarrow \nu_e$ dans l'exp\'erience NOMAD''}; Ph.D.
Thesis Universit\'e de Paris VII, Denis-Diderot, 1998.

A. Bueno \etal, NOMAD internal memo \# 98-016.

\bibitem{jura} 
ICARUS-CERN-Milano~Coll., CERN/SPSLC 96-58, SPSLC/P 304, December 1996;
J. P. Revol \etal, ICARUS-TM-97/01, 5 March 1997, unpublished.

\bibitem{boone} Updated information on the BOONE proposal can be found at 
{\it http://www.neutrino.lanl.gov/BooNE/}.

\bibitem{cernps} N.~Armenise \etal, CERN-SPSC/97-21, October 1997.

\bibitem{muonco} C.M. Ankenbrandt {\it et
al.}, Muon Collider Collab., ``Status Report'', 
see {\it http://www.cap.bnl.gov/mumu/}

\bibitem{geer} S. Geer, \Journal{\PRD}{57}{6989}{1998}.

\end{thebibliography}
\end{document}